# SPICES: spectro-polarimetric imaging and characterization of exoplanetary systems

## From planetary disks to nearby Super Earths

**Anthony Boccaletti · Jean Schneider · Wes Traub · Pierre-Olivier Lagage ·
Daphne Stam · Raffaele Gratton · John Trauger · Kerri Cahoy · Frans Snik ·
Pierre Baudoz · Raphael Galicher · Jean-Michel Reess · Dimitri Mawet ·
Jean-Charles Augereau · Jenny Patience · Marc Kuchner · Mark Wyatt ·
Eric Pantin · Anne-Lise Maire · Christophe Vérinaud · Samuel Ronayette ·
Didier Dubreuil · Michiel Min · Michiel Rodenhuis · Dino Mesa ·
Russ Belikov · Olivier Guyon · Motohide Tamura · Naoshi Murakami ·
Ingrid Mary Beerer · SPICES team**

The SPICES Team: **France**: A. Boccaletti (Obs. Paris), J. Schneider (Obs. Paris), P. Baudoz
(Obs. Paris), J.-M. Reess (Obs. Paris), R. Galicher (Obs. Paris), A.-L. Maire (Obs. Paris),
M. Mas (Obs. Paris), D. Rouan (Obs. Paris), G. Perrin (Obs. Paris), S. Lacour (Obs. Paris),
P. Thébault (Obs. Paris), N. Nguyen (Obs. Paris), L. Ibgui (Obs. Paris), F. Arenou
(Obs. Paris), J. F. Lestrade (Obs. Paris), M. N'Diaye (LAM), K. Dohlen (LAM), M. Ferrari
(LAM), E. Hugot (LAM), J.-L. Beuzit (IPAG), A.-M. Lagrange (IPAG), C. Vérinaud
(IPAG), P. Martinez (IPAG), M. Barthelemey (IPAG), J. Augereau (IPAG), S. Ronayette
(CEA), D. Dubreuil (CEA), P.-O. Lagage (CEA), E. Pantin (CEA), L. Mugnier
(ONERA) – **The Netherlands**: D. Stam (SRON), F. Snik (U. Utrech), M. Min (U. Utrech),
M. Rodenhuis (U. Utrech), C. Keller (U. Utrech) – **USA**: W. Traub (NASA JPL), J. Trauger
(NASA JPL), D. Mawet (NASA JPL), R. Belikov (NASA Ames), K. Cahoy (NASA Ames),
M. Marley (NASA Ames), M. Kuchner (GSFC), O. Guyon (U. Arizona/ Subaru), P. Kalas
(UC Berkeley), K. Stapelfeldt (NASA JPL), R. Brown (STScI), S. Kane (Caltech),
I. M. Beerer (MIT) – **Italy**: R. Gratton (Obs. Padova), S. Desidera (Obs. Padova), D. Mesa
(Obs. Padova), A. Sozzetti (Obs. Torino), A. Mura (IFSI Roma)– **Spain**: E. L. Martin
(INTA CSIC), H. Bouy (INTA CSIC) – **UK**: J. Patience (U. Exeter), M. Wyatt
(U. Cambridge), A. Allan (U. Exeter), R. King (U. Exeter), A. Vigan (U. Exeter),
L. Churcher (U. Cambridge) – **Switzerland**: S. Udry (Obs. Geneva) – **Japan**: M. Tamura
(NAOJ), N. Murakami (U. Hokkaido), T. Matsuo (NAOJ), J. Nishikawa
(NAOJ) – **Belgium**: C. Hanot (U. Liège) – **Germany**: S. Wolf (U. Kiel), L. Kaltenegger
(MPIA), H. Klahr (MPIA) – **Austria**: E. Pilat-Lohinger (Obs. Vienna).

A. Boccaletti (✉) · P. Baudoz · R. Galicher · J.-M. Reess · A.-L. Maire
LESIA, Paris Observatory, Meudon, France
e-mail: anthony.boccaletti@obspm.fr

J. Schneider
LUTH, Paris Observatory, Meudon, France

W. Traub · J. Trauger · D. Mawet
NASA-JPL, Pasadena, CA, USA

P.-O. Lagage · E. Pantin · S. Ronayette · D. Dubreuil
SAp/CEA, Saclay, France

**Abstract** SPICES (Spectro-Polarimetric Imaging and Characterization of Exoplanetary Systems) is a five-year M-class mission proposed to ESA Cosmic Vision. Its purpose is to image and characterize long-period extrasolar planets and circumstellar disks in the visible (450–900 nm) at a spectral resolution of about 40 using both spectroscopy and polarimetry. By 2020/2022, present and near-term instruments will have found several tens of planets that SPICES will be able to observe and study in detail. Equipped with a 1.5 m telescope, SPICES can preferentially access exoplanets located at several AUs (0.5–10 AU) from nearby stars (<25 pc) with masses ranging from a few Jupiter masses to Super Earths (∼2 Earth radii, ∼10 M$_\oplus$) as well as circumstellar disks as faint as a few times the zodiacal light in the Solar System.

**Keywords** Exoplanets · High contrast imaging

# 1 Introduction

SPICES (Spectro-Polarimetric Imaging and Characterization of Exoplanetary Systems) was proposed in 2010 to the ESA M3 Cosmic Vision call for missions for a launch in the 2020s. It is a direct imaging mission designed to achieve very high contrasts in order to characterize exoplanetary systems

D. Stam
SRON, Utrecht, The Netherlands

R. Gratton
INAF, Padova Observatory, Padova, Italy

K. Cahoy · R. Belikov
NASA Ames Research Center, Moffett Field, CA, USA

F. Snik
Sterrekundig Instituut Utrecht, Utrecht, The Netherlands

J.-C. Augereau · C. Vérinaud
IPAG, Grenoble, France

J. Patience
University of Exeter, Exeter, UK

M. Kuchner
NASA Goddard Space Flight Center, Greenbelt, MD, USA

M. Wyatt
University of Cambridge, Cambridge, UK

M. Min · M. Rodenhuis
University of Utrecht, Utrecht, The Netherlands

D. Mesa
Obs. Padova, Padova, Italy

previously identified by other instruments. SPICES is an evolution of the former SEE-COAST proposal [36] and belongs to the category of so-called small coronagraphic telescopes derived from the Terrestrial Planet Finder concept [26].

Recently, a crucial step has been accomplished with the first direct detections of both hot transiting giant planets [39, 40] and long-period warm young giants [16, 22, 25, 29]. Transit spectroscopy, though limited to transiting planets, was able to measure low-resolution spectra, both in emission and in transmission from the visible to mid-IR. Direct imaging has reached large contrasts mostly in the visible and near-IR. It has derived colors and first low-resolution spectra [4, 11]. Direct detections will be more frequent in the present decade with upcoming planet finders on 8 m telescopes from the ground and in space with James Webb Space Telescope (JWST), both in transit and direct imaging.

As a direct imaging mission, SPICES deals with exoplanetary systems that resemble the Solar System. Fed by a 1.5 m telescope, SPICES can preferentially access exoplanets located at several AUs (0.5–10 AU) from nearby stars (<25 pc) with masses ranging from a few Jupiter masses to Super Earths (~2 Earth radii, ~10 $M_\oplus$). SPICES, which will characterize planets by spectroscopy and polarimetry, has an important advantage over other exoplanet missions. Owing to its large field of view (10") and diffraction-limited imaging capability, SPICES can access planetary systems as a whole. In addition to the aforementioned planets (the main goal of the mission), much longer period planets (>10 AU) around young stars (found by planet finders on the ground) and circumstellar disks (from protoplanetary to debris disks) are accessible. Importantly, SPICES has the ability to detect exozodiacal light at the level of a few zodi, which is famously known to hamper the detection of Earth-like planets. This ability is therefore a tremendous advantage for future and increasingly ambitious direct detection programs.

This challenging goal clearly requires a specific instrumentation, which is not available on any other present or future facility and must be optimized for a very specific task. Typical star/planet brightness ratios are in the range of $10^8$ to $10^{10}$ at less than one arc-second. SPICES combines several techniques for high

O. Guyon
University of Arizona, Tucson, AZ, USA

O. Guyon
Subaru Telescope NAOJ, Hilo, HI, USA

M. Tamura
NAOJ, Tokyo, Japan

N. Murakami
Hokkaido University, Hokkaido, Japan

I. M. Beerer
Department of Aeronautics and Astronautics at MIT, Cambridge, MA, USA

contrast imaging in a single instrument designed to maximize the astrophysical return while reducing risks.

This paper summarizes the main objectives and characteristics of the mission as proposed to ESA. Section 2 describes the various types of objects SPICES will study together with their observable characteristics and gives a first estimate of the targets sample. The instrument concept is detailed in Sections 3 and 4 discusses the pointing issue. Some preliminary assessments of performances are shown in Section 5. Finally, general characteristics for the operations and the spacecraft are given in Sections 6 and 7. We note that a thorough estimation of performances will be presented in a forthcoming paper, while here we emphasize the science program and the technical aspects as presented in the Cosmic Vision proposal.

## 2 Science program

### 2.1 Observational approach

The exoplanet phenomenon is so diverse that it will require several missions and instruments from both space and ground to cover a broad range of parameters across several wavelength regimes. The primary objective of SPICES is to study the atmospheres and possibly surfaces of various type of planets: gaseous and iced giants as well as Super Earths. For that purpose, SPICES combines direct imaging with spectro-polarimetry in the visible (450–900 nm) and measures the total flux F and the linearly polarized fluxes Q and U, all as functions of the wavelength λ, for every pixel in an image. Combining spectroscopy and polarimetry has several advantages: first, the degree of polarization is sensitive to the properties of scattering particles and/or the reflecting surfaces; and second, polarimetry allows us to distinguish between reflected light by planets and light from stars which can be considered unpolarized. The ratio of the polarized fluxes to the total flux of a pixel determines the degree of polarization P and is given by:

$$P(\lambda) = [Q(\lambda)^2 + U(\lambda)^2]^{1/2} / F(\lambda) \qquad (1)$$

SPICES will be sensitive to exoplanets over a wide range of ages (a few Myr to a few Gyr), temperatures (warm to cold, respectively), and stellar types (from M to A stars), with arbitrary orbital inclinations and a wide range of orbital distances (from a fraction of an AU to several AU, depending on the stellar type and distance), and along large fractions of their orbits (depending on orbital inclination angles: at phase angles of $90° \pm 45°$). SPICES direct imaging and spectro-polarimetry will provide access to:

- the detection of unknown exoplanets in known planetary systems
- the orientation and stability of planetary orbits (yielding true masses of planets detected with the radial velocity (RV) method)
- the properties (density, microphysical characteristics) of exozodiacal dust

- the morphology of circumstellar dust and debris disks
- the composition of planetary atmospheres (absorbing gases)
- the presence and character of clouds/hazes
- the structure of planetary atmospheres (e.g. vertical distributions of gases, clouds, and hazes)
- the composition and structure of planetary surfaces (if present and visible through the atmosphere)
- the temporal variations in the composition and structure of planetary atmospheres and surfaces (due to seasons, eccentricity, chemical non-equilibrium, ...)

To accomplish these tasks, SPICES will have to face large contrasts (typically $10^8$ to $10^{10}$) at short angular separations (a fraction of an arcsecond). Table 1 gives the top level science requirements of the SPICES mission.

## 2.2 Characterization of cold exoplanets

Planets are colder when they are older and further away from their parent star. Cold planets emit relatively little thermal radiation and shine with reflected starlight at visible wavelengths. Flux and polarization spectra of this reflected starlight contain a wealth of information about an exoplanet's atmospheric composition, structure, and possibly surface (for rocky exoplanets). The fluxes depend on a planet's radius, orbital distance, and distance to the observer.

### 2.2.1 Gas and ice giants

Giant planets (both gaseous and iced) are definitely the main targets of SPICES since they will reflect more light than the emblematic but smaller Super Earths, so that a high level of atmospheric characterization can be achieved through both spectroscopy and polarimetry. We anticipate that 1–5 AU orbits around nearby stars will be perfectly suited to SPICES (see Section 5).

Both flux and polarization spectra have an overall shape, the continuum, upon which narrower spectral features are superimposed. The level and shape of the continuum are mostly determined by scattering and absorption by gases (Rayleigh scattering), clouds, and aerosols. Most of the narrow spectral features in total flux F and polarization P spectra arise from absorption of light by atmospheric gases. The depth and width of the spectral features yield information about overall atmospheric pressure and the existence and altitude of clouds and differ from one planet to the other (sometimes significantly) as shown in Figs. 1 and 2. With SPICES' spectral resolution of about 40, we will be able to identify various types of gases.

**Table 1** Scientific requirements for SPICES

| Quantity | Science requirement | Instrument requirement |
|---|---|---|
| Star-to-planet contrast | $1R_J$ at 3 AU with 50% albedo | $<5.10^9$ |
| Angular resolution | Planets at $a > 1$ AU at 10 pc | 100 mas at 600 nm |
| FOV inner part of disks | (10–50 AU) | >5" (goal 10") |
| F accuracy | Properties of atmospheres | <5% |
| P accuracy | Properties of atmospheres | <5% |
| Spectral range | Main molecules, Rayleigh | 450–900 nm |
| Spectral resolving power | Main molecules, Rayleigh | R = 40 |

**Fig. 1** Spectra of the gas and ice giant planets in our solar system from [23] smoothed to SPICES resolution, R = 40

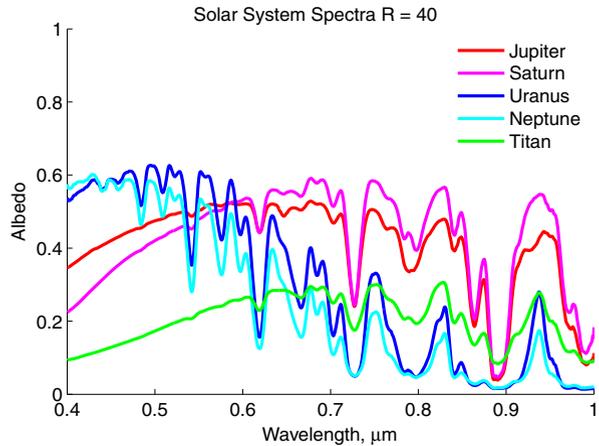

The spectral signatures from cloud and aerosol layers depend on the particle properties, the number of particles and their spatial distribution, the illumination and viewing geometries, and the wavelength. These signatures show up differently in flux and polarization spectra, and the greatest insights can be gained by using both: total flux spectra are sensitive to the amount of scattering (scattering cross sections, particle number densities, scattering phase functions), while polarization spectra are more sensitive to the microphysics of the particles in the atmospheric layer where the last scattering takes place.

**Fig. 2** Albedo of model Jupiter and Neptune planets [13] at separations of 0.8 AU (*red* and *dark blue*, too warm for clouds) and 2 AU (*green* and *cyan*, just cold enough for water clouds) smoothed to SPICES resolution, R = 40

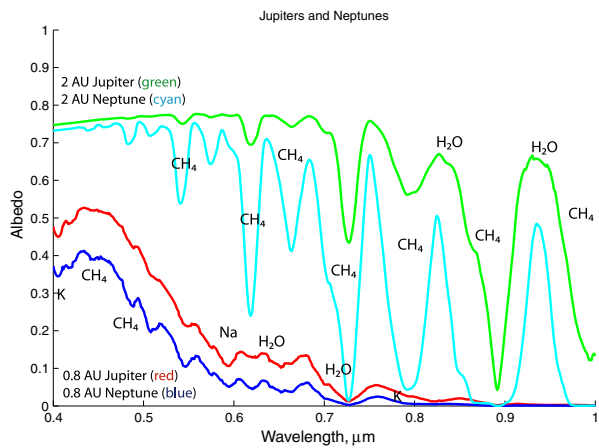

For all of the Solar System giant planets, at redder wavelengths we can see deeper into their atmospheres because Rayleigh scattering and haze absorption decrease. Figure 1 shows that the geometric albedos of Jupiter and Saturn are large at longer wavelengths because these planets have thick clouds relatively high in their atmospheres that efficiently reflect sunlight back to space. Shorter (bluer) wavelengths are absorbed by photochemical hazes instead of Rayleigh scattered. The geometric albedos of Uranus and Neptune are much lower but clearly show Rayleigh scattering at the shortest wavelengths, because for these colder planets, most of the cloud layers are located deeper in the atmosphere, which contain significantly more methane gas. Methane gas has many useful, measurable absorption bands of various strengths across SPICES spectral range, as shown in Figs. 2 and 3.

SPICES will constrain the evolutionary architectures of extrasolar planetary systems because it will be able to tell the difference between ice and gas giant planets, even with a low resolution of R = 40. The evolutionary history ultimately determines the composition and structure of gas and ice giants, driving factors that affect their spectra, such as gas mixing ratios and clouds [13]. For example, giant exoplanets that are somewhat warmer than Jupiter should show $H_2O$ water vapor absorption in the red [13]. SPICES will also be able to characterize the atmospheric temperature and composition of a far greater variety of giant planets than found in our own solar system. Extrasolar planets with ages of 200 Myr to 10 Gyr and masses of 1–5 Jupiters will have effective temperatures ranging from 500 K to below 100 K. As shown in Fig. 2, warmer cloud-free atmospheres reflect less and thus have darker spectra, especially at longer (redder) wavelengths. In the absence of clouds, sodium Na (0.59 μm) and potassium K (0.78 μm) are visible. Clouds increase

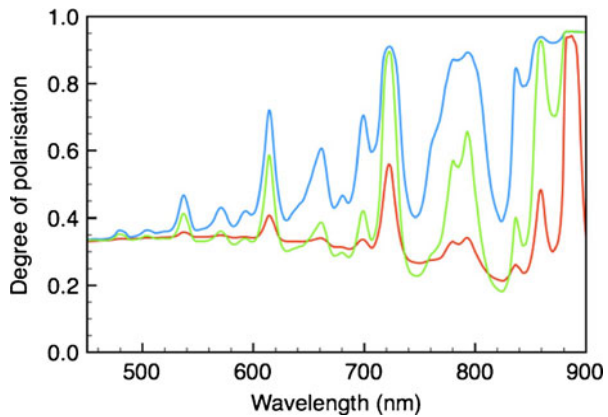

**Fig. 3** Simulated polarization P spectra (R = 40) of a Jupiter-like planet with a *thick cloud layer* (*red*), and two Neptune-like planets (*blue* and *green*), with much more methane absorption and deep lying clouds at two altitudes

the amount of reflected light (albedo) and also increase the contrast between continuum and adjoining strongly absorbing spectral regions.

As temperature decreases for older or further-separated planets, the first clouds to appear in the skies are $H_2O$ water clouds. If temperatures get even colder the altitude of the $H_2O$ water clouds also gets lower and deeper into the atmosphere (since atmospheric temperature falls with altitude). Next, high-altitude $NH_3$ ammonia clouds will form (with spectral features at 0.55, 0.65, and 0.79 μm that are not shown in these plots), followed by the formation of $CH_4$ methane clouds (strong features in 0.7–0.9 μm as shown in Fig. 2). Thus, the water and ammonia clouds that will likely be found in the atmospheres of planets in SPICES detection range will indicate atmospheric temperature and composition [13].

In addition to being able to differentiate between gas and ice giant planets, SPICES will also be able to differentiate gas giant planets from brown dwarfs. This ability is extremely useful for validating or ruling out planet candidates observed by previous direct detections, where mass is derived photometrically (such as JWST or planet-finders on the ground).

All giant planets in our Solar System have rings. SPICES cannot spatially resolve rings, but they reveal their presence by reflecting starlight and by casting their shadow on the planet. The traces of rings in the planets flux and polarization phase functions depend on ring properties such as radius, optical thickness, particle microphysics, and the geometry with respect to the planet and its orbit [2]. The high photometric stability of a space mission like SPICES allows the ring presence to be revealed.

### 2.2.2 Nearby super earths

Super Earths are very interesting objects but obviously fainter than giants and necessarily rare in the solar neighborhood (due to RV detection bias). SPICES will be able to detect large terrestrial planets (>2 Earth radii) with Earth-like albedos around a few of the nearest stars (<5 pc). The largest terrestrial planets in the Solar System (Venus, Earth, Mars) differ strongly in composition, surface pressure, cloud coverage, and surface type at visible wavelengths. They each have unique flux F and polarization P spectra. In addition, Super Earths will have spectra that largely contain information about their clouds and surfaces. The Chappuis band of ozone $O_3$ (590 ± 50 nm) is a prominent broad feature as well as the oxygen $O_2$ A band (760 nm ± 10 nm) and various water $H_2O$ bands. In favorable circumstances, SPICES could detect the most obvious spectral features (Section 5) and polarization effects providing the degree of polarization is large like for Titan (see Fig. 4).

SPICES will see the lowest possible scattering surface of a planets atmosphere. This surface might be the top of clouds or the solid/liquid surface of a few nearby Super Earths. Sensing the actual surface will only be possible if the planets atmosphere is optically thin, like that of Mars, or if the cloud coverage is sparse enough, such as on Earth. A particularly interesting surface type to search for would be liquid water. For that, SPICES could take advantage of its polarization spectra capability, since oceans generally appear dark and have very low albedos and thus may not be detectable in flux F spectra. SPICES can make the difference between various type of surfaces (clouds/forests/oceans) with very strong color effects in the visible. Observations of a planet at different locations in its orbit could reveal seasonal effects provided they are strong, as they might be for a planet in a highly elliptical orbit or with a large obliquity.

**Fig. 4** The measured polarization of planets in the solar system [35]. Titan-like planets can have very large degree of polarization

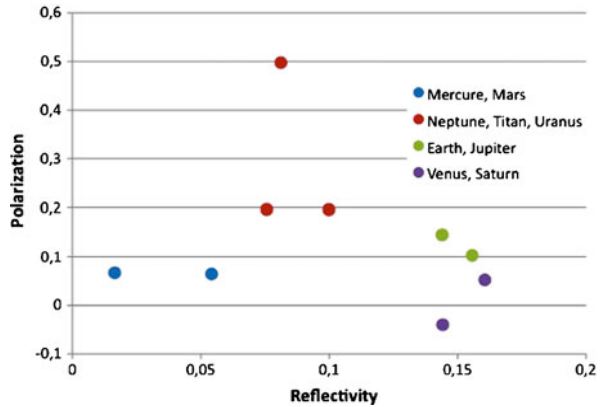

## 2.3 Characterization of young and warm exoplanets

Planets around young stars are young and warm, and their thermal emission makes them relatively bright at visible wavelengths. Any planets discovered at longer wavelengths with the upcoming next-generation adaptive optics exoplanet instruments, SPHERE (Spectro Polarimetric High contrast Exoplanet REsearch [8]), GPI (Gemini Planet Imager [28]), and HiCIAO (High Contrast Instrument for the Subaru next generation Adaptive Optics [41]) will also be detectable with SPICES (Fig. 5). The emission from these warm planets will still be dominant in the red part (I band) while they will be much fainter (and possibly out of reach of SPICES) at shorter wavelengths owing to their large orbital distances (some tens of AU). It is a unique opportunity to combine visible and near IR spectra to learn more about gaseous absorptions, temperatures and clouds. In addition, depending on the stellar spectral type, SPICES can extend the detection limit down to one Jupiter mass and even one Saturn mass. Due to symmetry, the thermally emitted radiation of young exoplanets will be mostly unpolarized when a planet is observed as a point source. If polarized planetary thermal radiation is in fact observed, it would indicate a horizontally non-homogeneous atmosphere, e.g. an atmosphere with patches of clouds, again complementing near IR ground-based observations.

**Fig. 5** Absolute I band magnitude of young planets (assuming BD-COND model from [3]) compared to the expected detection limits of SPICES for early and late type stars. SPICES will overcome the 8-m class planet finders by several orders of magnitudes (the 8-m class level is effectively shown by the currently detected planets)

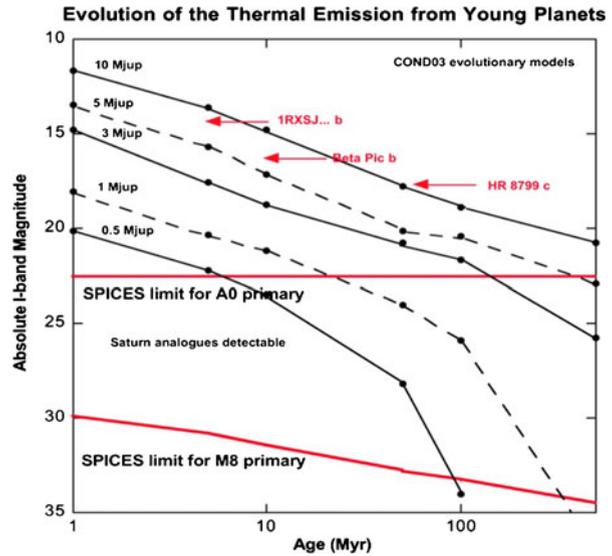

## 2.4 Characterization of circumstellar disks

SPICES will be able to study a large range of circumstellar disks from protoplanetary to exozodii representing different stages of disk evolution (in proportion with the amount of dust).

Up to now, very few protoplanetary disks have been angularly resolved by direct imaging in their inner (planetary) regions (1–50 AU). The SPICES observatory offers a unique point-source rejection which allows for the first time ever, the possibility to study the geometry and the dust properties of the inner regions of proto-planetary disks with emphasis on the planetary formation. The information from SPICES on the disks scattered light will be entirely complementary to mid-IR (JWST) and sub-mm (ALMA: Atacama Large Millimeter Array) data, since different wavelengths probe dust in different size ranges where different physical processes operate. Polarimetry (degree and direction) will be able to lift important degeneracies in determining the 3D structure of the disks.

The majority of the photometrically-detected debris disks (by Spitzer for instance) are predicted to be ring-like at radii in the 0.1–3" range accessible to SPICES. Recent high resolution images obtained on 8-m class telescopes illustrate this characteristic [12, 42]. Planetary perturbations are known to cause disk structures like offsets, warps, clumps, and spirals, as confirmed by [22, 25]. Therefore, high contrast images can also be used to identify signposts of new planets and provide constraints on the planets' masses, orbits and even evolutionary history.

Exozodiacal dust generated by sublimation of comets and collisions among asteroids can be both friend and foe. SPICES can perform a census of zodiacal

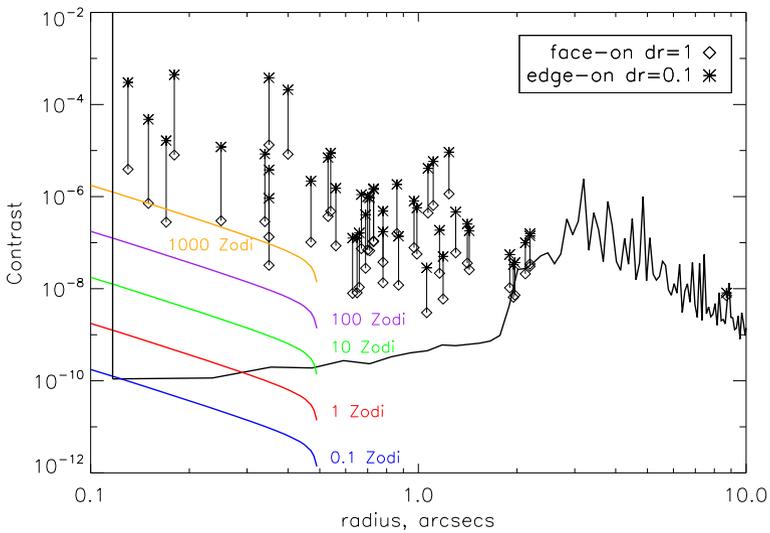

**Fig. 6** Simulating SPICES observations of the 46-A star debris disks that were detected at both 24 and 70μm with Spitzer shows that all lie above the contrast limit of SPICES (given here at 1 sigma), regardless of disk orientation. Levels of Zodiacal light are shown for 0.1–1000 Zodi. The level of detection corresponds to a sensitivity of a few zodi

clouds around nearby stars that would be important targets for future missions for direct imaging of exo-Earths, therefore surpassing any currently proposed mission focusing on exo-zodi (Fig. 6) with the ability to directly recover extended structure geometry independently from any models.

## 2.5 Targets

The SPICES target sample relies mainly on previously detected objects: known planets (M < 25 M$_J$) and disks. The mission will be defined with these observational constraints, although a survey-like approach is also envisaged for some nearby stars but as a second priority (with no strong drivers on the instrumental concept).[1] Current and near-term detection programs will provide appropriate targets for SPICES, although it is difficult to predict the exact number. However, we can make estimations based on the current performances and the progression rate of discoveries. Our goal in this section is (1) to show that some known planets are already achievable with SPICES and (2) to demonstrate that a significant number of objects will be available by the 2020s, actually more than SPICES could ever observe. Five classes of targets are relevant for SPICES. Overall, there will be at minimum one hundred targets suitable for characterization with SPICES and certainly some hundred objects to search for. We estimated that, given the exposure time needed for characterization and surveys, our program requires a five-year mission.

---

[1]This in particular allows us to reduce the number of deformable mirrors (see Section 3.5).

**Fig. 7** Planets known from RV surveys and potentially observable with SPICES as of December 2010. *Red dots* are for planets angularly separable at the shortest wavelength (450 nm) and *blue dots* are for 900 nm

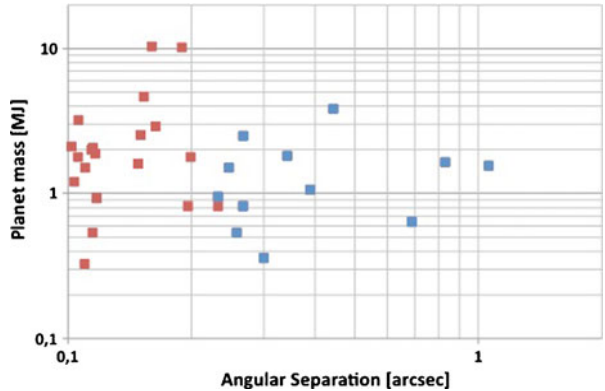

### 2.5.1 Planets from radial velocity

Among the nearly 470 planets known from RV,[2] more than 145 stars have planets more massive than 1 $M_J$ and at separation larger than 1 AU (according to http://exoplanet.eu). Only a fraction of these planets will be observable with SPICES depending on the projected angular separations. In the current database, almost 30 objects match the SPICES fundamental limitations (Inner Working Angle ∼ 0.1" @ 0.45 μm) with masses larger than 0.5 $M_J$ (Fig. 7). If we assume that about 3.3% of stars have planets in the 3–6 AU range [46], this translates to about 60 giants observable with SPICES within 20 pc.

As for telluric planets, the current precision of HARPS (High Accuracy Radial velocity Planetary Search project) has already allowed for the iden- tification of some 20 candidates (unpublished as of yet) in the range of 4–10 $M_\oplus$ at short periods (0.05–0.5 AU). There is no doubt that both HARPS then ESPRESSO (successor of HARPS at the VLT) will push these numbers of detection further in the appropriate range for SPICES. Howard et al. [21] found that 6.5% of solar-type stars have close-in planets of 10–30 $M_\oplus$ which means a few units of Super Earths observable with SPICES. Overall, if the rate of discovery follows the current trend, we can reasonably predict that on the order of 100 RV long-period (>1 AU) planets will be observable with SPICES including 5–10 Super Earths.

Furthermore, it is clearly demonstrated that the planetary systems discov- ered with RV are usually packed, which means that longer periods have higher probability when close-in planets already exist [27]. This potentially extends the target sample of SPICES to hundreds of objects for a survey-like approach. It is a very strong argument in favor of a direct imaging mission.

---

[2]as of December 2010

### 2.5.2 Planets from astrometry

So far, very few planets are confirmed and none have been discovered by astrometry, but it is anticipated that the next European astrometric mission GAIA (Global Astrometric Interferometer for Astrophysics) will open up a new discovery space (unbiased spectral types, all sky, volume limited) towards long periods, perfectly suited for SPICES. GAIA (launch in 2012) should provide targets timely for SPICES and more importantly accurate masses. Given some assumptions detailed in [14] about planet probability, GAIA can observe hundreds of stars in the 0–20 pc distance bin ($6 < V < 13$) and should detect a dozen giants (more massive than Saturn) with 0.5–4.5 AU semi-major axis, therefore suitable to SPICES.

### 2.5.3 Planets from direct imaging

Direct imaging has now started to detect several long-period planets around relatively young stars [22, 25, 29] and even planet in formation [24]. About 12 such objects are known as of today and the future instruments will yield tens of new young objects around a large variety of spectral types (A to M) for nearby stars ($<$50–100 pc). It is anticipated that early type stars will provide more detections while they may form more massive planets. For now, most of these planet candidates are orbiting relatively far ($\beta$ Pic b being the closest at 8 AU) but could be warm enough to be bright at visible wavelengths. The sensitivity of SPHERE, GPI, and JWST will be sufficient to detect planets at typically $>$5 AUs, which will make good targets for SPICES (although certainly fainter than those found by RV in reflected light). SPHERE also has the capability to detect Jupiter-like planets at 0.5–1 AU for very bright nearby stars in the visible ($\alpha$ Cen is a good candidate for instance). We can estimate that after five years of operations (so before 2020), possibly tens of planets detected by 8-m class telescopes will be accessible to SPICES, improving the level of characterization.

### 2.5.4 Planets from transit

The space mission PLATO (PLAnetary Transits and Oscillations of stars) proposed to Cosmic Vision in 2007 could observe a few bright and nearby stars: 90 targets with $V < 6$ and 1,350 targets with $V < 8$, according to [15]. Assuming a 1% transiting probability this translates into about 1 and 10 potential targets in these two magnitude bins. Although marginal, PLATO may add a few units to the target sample of SPICES especially in the Super Earth mass regime. More importantly, these planets will have their mass, radius, and then density measured, so that SPICES can directly derive albedos. Similar to RV, the presence of planets in close-in orbits is a good criterion for searches at longer periods. In addition, for these planets, SPICES can make combined light spectroscopy observations during transits.

**Table 2** Evaluation of mission duration

|  | Characterization | Disks | Survey |
|---|---|---|---|
| Number of targets | 100 | 100 | 100 |
| Number of targets to revisit | 20 | 20 | 0 |
| Number of visits | 5 | 3 | 1 |
| Individual integration time (hours) | 100 | 30 | 50 |
| Total integration time (years) | 2.0 | 0.50 | 0.6 |
| Total including overhead, calib/pointing (years) | 2.8 | 0.65 | 0.8 |

### 2.5.5 Disks

There are already several tens of known protoplanetary and debris disks observed by direct imaging with HST and from the ground that are mostly around massive stars. SPITZER also provided many targets with IR excesses around lower mass stars, not yet resolved, which SPICES will be able to reveal better than HST.

### 2.5.6 Mission duration

According to the previous subsections about 80 to 100 targets would constitute the core sample of SPICES for deep spectro-polarimetric characterization. Similarly, about 100 known protoplanetary and debris disks are observable with SPICES. Then, an additional program, a survey of the closest and brightest stars can be accomplished (there are 120 stars of V < 5 within 20 pc). Performances and signal to noise ratios are analyzed in Section 5, and although the exposure is a strong function of stellar and planetary parameters (Sp, distance, separation), we here assume a generic and conservative value of 100 h. Table 2 gives a crude and first estimate of a design reference mission to carry out in a five-year mission.

## 3 Instrumental concept

### 3.1 Orbit requirements

For the purpose of thermal stability (high dimensional stability is required for the telescope and instrument optics), target accessibility (avoidance of occultation of the sky by the Earth), and high data rate for the full mission, SPICES will be on an orbit around the Sun-Earth L2 Lagrangian point. There are several potential L2-orbit families (halo, Lissajous, ...). The pointing strategy is analogous to that used on the Herschel mission. It has two degrees of freedom: some 30° of tilt around the plane normal to the Earth-Sun direction and a rotation around the Sun direction. As a matter of fact, the conjunction of Earths orbital movement and pointing strategy allows any target to be accessible at least twice a year, and visibility duration for a target varies with its declination. To reach L2, the proposed launcher is a Soyuz–Fregat launched from Kourou. The spacecraft, with an estimated launch mass around 1.5 tons, will fit in the fairing.

## 3.2 General overview of the payload

The payload concept is designed to fulfill the science requirements given in Section 2. The general problem to directly image extrasolar planets near bright stars is well known and emblematic techniques like coronagraphy and wavefront control have been developed over the last 15 years to enable high-contrast. Although potential solutions are numerous, we propose a single conceptual design for the payload. The design of SPICES results from a trade-off between science requirements (level of characterization) and performances (stability, contrast, ...).

SPICES is an off-axis 1.5 m telescope consisting of a coronagraphic system combined with wavefront correction and feeding an integral spectro-polarimeter to cover the 450–900 nm band. The optical quality of the telescope is not drastic (typically 10 nm rms at mid frequencies on the primary). The wavefront control is achieved with the Electric Field Conjugation [20], a specific algorithm to measure the aberrations, and a deformable mirror (DM) to correct the wavefront, providing a quality/stability on the order of tens of picometers. Our goal is to implement a more promising technique, the Self Coherent Camera [18], which for now has a lower Technology Readiness Level (TRL) but will allow a better discrimination of planets and speckles [5]. We chose the Vector Vortex Coronagraph (VVC) [32], a derivation of the phase mask concept which can be made potentially achromatic on a ~50% bandwidth. The backend instrument is a micro-lenses based integral field spectrograph (IFS) [1] similar to those developed now on the ground for SPHERE and GPI. Polarimetry is intimately implemented in the design by using a rotating half-wave retarder as a modulator and a polarizing beam-splitter cube as an analyzer.

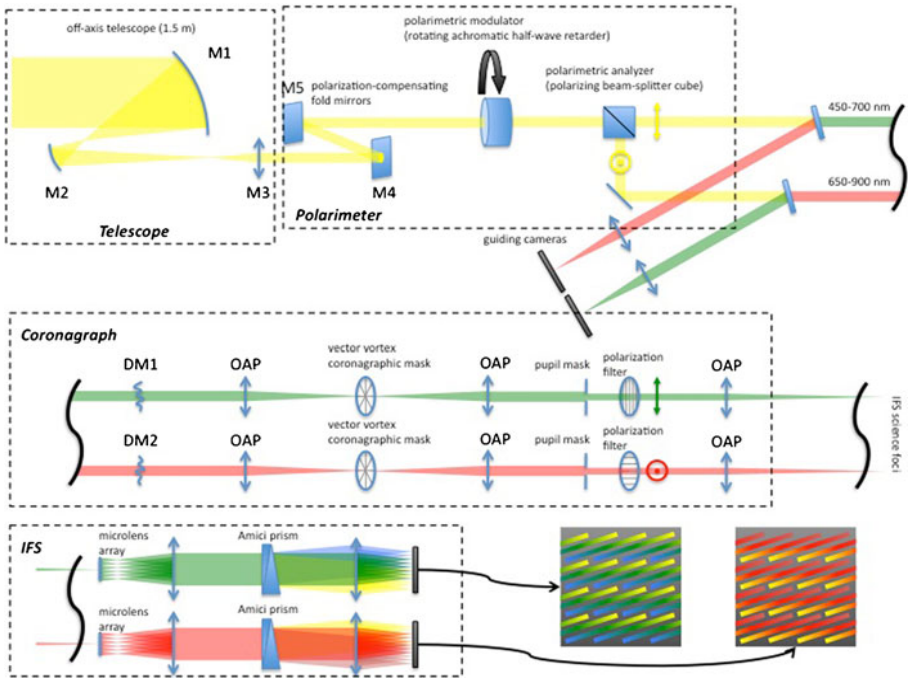

**Fig. 8** Conceptual design of the SPICES payload showing the main blocks: telescope, polarimeter, coronagraph and IFS. Only the main optics are shown here for sake of clarity (see Sections 3.2 and 3.3 for a description)

Figure 8 illustrates the conceptual design of the instrument. A polarimetric beam splitter splits the beam into two channels. One polarization is assigned to one channel (for instance vertical polarization at the top, horizontal polarization at the bottom, as in Fig. 8). A modulator is used upstream (between the telescope and the beam splitter) to select the linear polarization direction on the sky that is analyzed by the polarizer in each channel: ±Q (horizontal and vertical linear polarization) and ±U (linear polarization at ±45°). This furnishes a complete measurement of linear polarization of the incoming light (Fig. 9).

In addition, each channel is specialized for a spectral bandwidth, 450–700 nm and 650–900 nm (with an overlap of 50 nm for calibration purposes). The two-channel concept allows flexibility to optimize the wavefront measurement and correction, the coronagraph achromaticity, the IFS design, and then the whole chromaticity of the instrument. This solution also reduces the number of mechanisms from two to three (the modulator at the entrance of the instrument and the tip/tilt mounts) and gives redundancies to avoid single point failure for critical elements like DMs. The conceptual design also shows that half of the light is sent to the two science channels and the other half is not wasted but used for internal guiding (see Section 4).

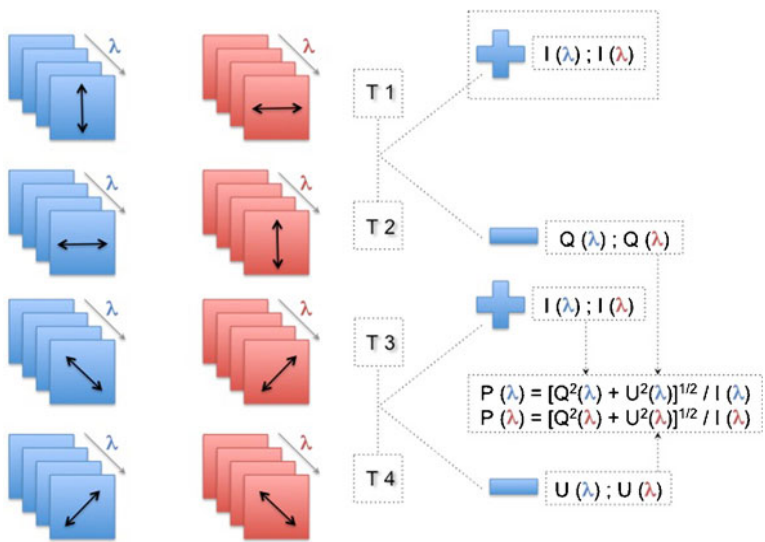

**Fig. 9** Illustration of the SPICES measurement concept. Data cubes (x,y,λ) are reconstructed for the *blue* and *red* IFS channels. Four sequences are obtained at different time T1, T2, T3, T4. At a given time the polarization directions are orthogonal between the *blue* and the *red* channel. The direct sum of T1 and T2 data gives the intensity spectrum while two more sets of data (with 45° and 135° orientations) are needed to build Stokes U and Q vectors and hence the degree of polarization

With this concept, a typical observing sequence on an exoplanetary system will require four observations to collect all the spectral and polarimetric information (Fig. 9). The whole optical design, telescope + instrument, needs to be specified with a good optical quality. Even if active correction is implemented with deformable mirrors, the total wavefront error budget must be low, a few tens of nanometers (goal: 20 nm rms). Similarly, the performance of SPICES is contingent on the ability to accurately measure the aberrations. Spectral decorrelation due to out-of-pupil optics must then be controlled. This aspect is an important requirement in the optimization of the optical design. The high level specifications of the instrument are given in Table 3.

## 3.3 Optical layout

SPICES includes an unobstructed afocal telescope and a coronagraphic device feeding an integral spectro-polarimeter. The three mirrors of the 1.5 m afocal telescope (2 Off Axis Parabola + 1 pure elliptical concave mirror) provide a 64 mm collimated beam to match the DM size (Section 3.5). The off-axis primary mirror (PM) of SPICES can be directly inherited from the GAIA mirrors, which have very similar specifications and achieve a Wave Front Error (WFE) of 7–8 nm rms (on the surface). GAIA is also a cold telescope (200 K) with a high degree of stability (at a level of 15 pm), similar to what is envisaged for SPICES. Then two fold mirrors (M4 + M5) are used to balance the polarization induced by the telescope before entering the polarimetric modulator and the polarizing beam-splitter cube. The beam is then separated into two channels dedicated to the 450–700 nm and 650–900 nm bands. On each arm, a DM (including the tip-tilt function in its mounting) is located in the close pupil image plane, then a first stage provides an image plane for the Vortex (F/60) with a WFE (RMS) = $\lambda/100$ on the edge and $\lambda/2500$ on center (at 450 nm). A following second stage provides a 10 mm pupil image for the coronagraphic diaphragm, a broadband filter and a polarization filter, and finally a third camera stage provides a telecentric image plane (F/290) to feed the IFS with a WFE (RMS) = $\lambda/150$ (at 450 nm). Based on OAP and fold mirrors, the whole optical design is very flexible for further optimization with respect to chromaticity (Fresnel propagation) and the mechanical implementation of the DM, the vortex, the SCC option or the exact IFS locations. Figure 10 shows the proposed optical layout.

**Table 3** Overall requirements for the payload

| Parameters | Values |
| --- | --- |
| Telescope diameter (D) | 1.5 m |
| Bandwidths | 450–900 nm (goal: 400–950 nm) |
| Equivalent spatial resolution | 62–120 mas (goal: 55–127 mas) |
| Blue channel/red channel | 450–700 nm/650–900 nm |
| Observable Stokes parameters | I, Q, U |
| Contrast at 2 lambda/D | a few $10^{-9}$ |
| Contrast at 4 lambda/D | a few $10^{-10}$ |
| Deformable mirror, nb. of elements | $64 \times 64$ actuators |
| FOV corrected blue/red channel | 6"/8" |
| FOV imaged blue/red channel | 9"/12" |
| WFE static | 20 nm rms |
| WFE DM | 10 pm rms |
| Final pointing accuracy | 0.5 mas (goal 0.1 mas) |
| Polarimetric sensitivity | $10^{-3}$ |

3.4 Coronagraph

Coronagraphy is essential in the SPICES design to get rid of the stellar photon which in turn attenuates the impact of the photon noise. The main science requirements of SPICES translate into high contrast $(10^8 - 10^9)$ and small Inner Working Angle (IWA, $< 2\lambda/D$). Phase masks gather such advantages [34] and have been intensively studied and implemented in the lab [6, 9, 30] and on the sky [10, 31] in different forms. The Vortex Coronagraph (OVC, VVC)

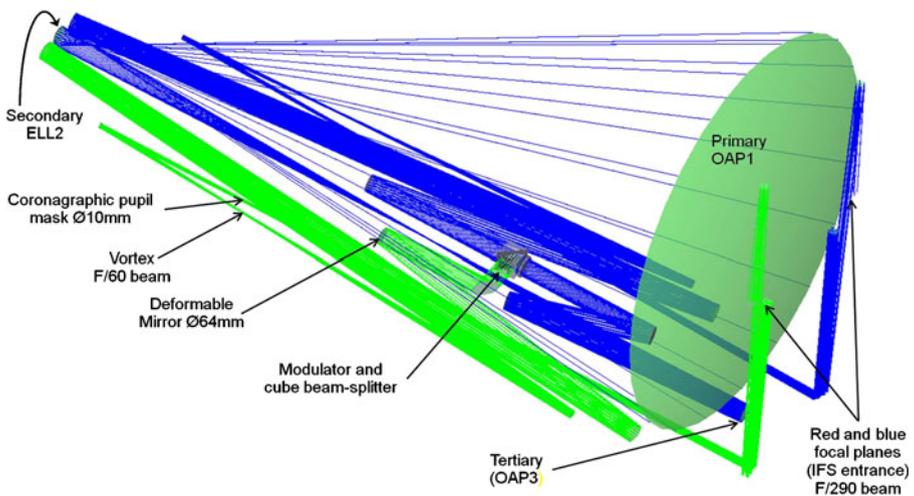

**Fig. 10** General optical layout showing the implementation of the main components. Most optics are located below the M1–M2 segment while the IFSs are at the *back* of the PM

is an evolution of this concept [32], which in addition improves the discovery space. The advantages of an optical vortex coronagraph are that it can provide imaging very close to the star ($< 2\lambda$/D) and high throughput ($>90\%$). The first generation of optical charge 4 VVC (TRL 6) has demonstrated $10^{-7}$ contrast levels over 10% bandwidth using wavefront correction [32] and is now pushing towards higher contrasts and larger bandwidths. The vectorial nature of the VVC and its intrinsic polarization modification property allows the complete decoupling of the chromatic leakage from the main vortex term and any transmitted off-axis features (companion and/or disks). Very recently, a vector vortex mask made of photonic crystal has been developed and achieves contrast levels of $5.10^{-6}$ and $1.10^{-6}$ at 5 $\lambda$/D for wavelengths of 532 and 633 nm, respectively (Murakami et al., in preparation). The demonstration of an achromatic VVC, which meets the requirement of SPICES is now at TRL 3.

3.5 Wavefront sensing and correction

Several techniques exist to estimate phase and amplitude aberrations of the wavefront. The Electric Field Conjugation (EFC, baseline) uses temporal modulation whereas the Self-Coherent Camera (SCC, goal) uses spatial modulation.

EFC [20] is a wavefront control technique that is particularly well-suited for high contrast coronagraphy in space. This is because it uses the science camera to sense aberrations, so that there are no non-common path errors; it directly removes stellar speckle noise in the region of interest regardless of where that speckle noise came from; and it uses the DM that is already in the system to provide the diversity needed for estimation of aberrations in the focal plane. EFC has been successfully demonstrated on many testbeds and with many coronagraphs, as well as in monochromatic and broadband light. For example, record-breaking contrasts have been achieved by some members of this proposal at the NASA JPL High Contrast Imaging Testbed [43] as well as the NASA Ames Coronagraph Testbed [7].

As for the SCC, the easier way to implement it at almost no cost for the optical design is to add a small hole (reference beam) in the Lyot stop plane of the coronagraph. The last coronagraph optics make the classical coronagraphic beam and the reference beam interfere in a Fizeau scheme on the detector. The speckles of the coronagraph beam are spatially modulated by fringes. As the beams from the star and the companion are not coherent, only the speckles are modulated (the companion image is not). The SCC is intrinsically chromatic and in SPICES this issue is solved with the use of the IFS (same for the EFC). Currently, the SCC is being implemented in a high contrast test-bench and the first measurements of phase and correction are being obtained (TRL 3). Although the maturity is low compared to the EFC, it has the advantage to avoid temporal modulation and to provide additional criteria to get rid of the speckles further than EFC. More details and performance are given in [18, 19].

Together with the wavefront sensing methods described above, SPICES uses a deformable mirror (DM) for wavefront control. The DM manipulates both the amplitude and phase of the wavefront simultaneously to create the high-contrast coronagraph dark field. Wavefront control at the $10^{-9}$ level has been demonstrated in JPL laboratory experiments for the past five years [45]. SPICES baselines proven deformable mirror technology manufactured by Northop Grumman Xinetics. Specifically, the DM is comprised of a fused silica mirror facesheet driven by an array of electroceramic actuators. Protoflight qualification of a flight-configured $48 \times 48$ DM is now in progress at JPL, validating the technology. With years of performance in a space-simulating environment, and following successful preliminary vibration tests at JPL, the DM technology is currently at TRL5+ with expectations of reaching TRL6 in environmental tests soon.

## 3.6 Polarimetry

The polarimetric capability of SPICES is essential to fulfill its science requirements and therefore constitutes an integral part of its design. Any modern astronomical polarimeter consists of a modulator and an analyzer.

To measure the observables for SPICES, a modulator is positioned in the beam upstream from the analyzer after the first five mirrors, in a collimated beam. It can either be a double Fresnel rhomb or a so-called "superachromatic" half-wave plate constructed from three stacks of quartz and $MgF_2$ that serves the full wavelength range of 450–900 nm. Essentially, such a modulator rotates a direction of polarization on the sky to the one direction that the analyzer filters. With a stepwise rotation of 22.5° increments, the intensities that the polarimeter measures are $(I + Q)/2$, $(I + U)/2$, $(I - Q)/2$, and $(I - U)/2$. Thus, the complete information on linear polarization is obtained with a single optical/detector train.

The baseline concept of SPICES presented in Fig. 8 uses a polarizing beam-splitter as a polarimetric analyzer. For the baseline configuration we selected a cube beam-splitter, because of its achromatic splitting angle. This implementation is a single-beam system, as the different intensity measurements are obtained sequentially. This drives the requirement on the typical time-scales of pointing and thermal variations to be much larger than the modulation cycle time. Although dual-beams are required on ground-based instruments, the sensitivity of the single-beam systems in SPICES will likely be better than $10^{-3}$, which is more than enough for most targets.

| | Blue channel | Red channel |
|---|---|---|
| Wavelength range (nm) | 450–700 | 650–900 |
| FOV (arcsec) | 9.24 | 13.3 |
| Resolution | 42 | 55 |
| IFU pitch ($\mu$m) | 70 | 70 |
| N lenslets (side) | 299 | 298 |
| F/number at the entrance | 289.5 | 200.4 |
| Spectra length (pix.) | 36.52 | 35.48 |
| Approx. IFS size (mm) | 830 | 830 |

Table 4 IFS main parameters for the two channels assuming $4 \times 4$ k detectors

**Table 5** Characteristics of science detectors

| Parameter | Value | Comments |
|---|---|---|
| Dimension | $4,000 \times 4,000$ | |
| Pixel pitch | 15 μm | |
| Quantification | 12 bits | Goal 16 bits |
| Full well capacity | 300 ke- | |
| Readout frequency | <0.1 Hz | 19 Mbits/s for full frame at 0.1 Hz |
| Readout mode | Full frame | |
| Readout noise | <5 e- | Goal 1e- |
| Dark current | <5 e-/h | Need a detector temperature <170 K |
| Flatness | <20 μm | |

Polarizing beam-splitters and rotating polarization modulators are currently employed in various solar observing satellites that have polarimetric functionality.

## 3.7 Integral field spectrograph

The proposed Integral Field Spectrograph (IFS) is based on the lenslet BIGRE concept [1] in manufacturing for SPHERE [17]. Each lenslet of an array samples a portion of the field of view and provides a virtual slit at the input of a dispersing spectrograph, made up of the usual components (collimator, dispersing element, and camera). The dispersing elements are direct vision

**Table 6** Summary of TRLs for the main components as of year 2010

| | TRL | Rational |
|---|---|---|
| Telescope | | |
| – Primary mirror | 7 | Process qualified on similar space missions (GAIA) |
| – Secondary mirror | 7 | Process qualified on similar space missions (GAIA) |
| – Baffles | 7 | Process qualified on similar space missions (GAIA) |
| – Structure | 7 | Process qualified on similar space missions (GAIA) |
| Science payload | | |
| – Rotating polarization modulator | 9 | Hinode Solar Optical Telescope |
| – Polarizer cube beam splitter | 9 | Hinode Solar Optical Telescope |
| – Filters | 5 | Process qualified on similar space missions (MIRI/JWST) |
| – Deformable mirrors | 5 | Space qualifications at JPL |
| – Coronagraph | 3 | Process verified in lab environment |
| – IFS | 5 | Concept used on ground environment (SPHERE) |
| – Relay optics | 5 | Process qualified on similar space missions |
| – WF sensing | 5 | Baseline concept qualified at JPL |
| – Detectors | 6 | E2V CCD 231 − 84 back illuminated |
| – Mechanical structures | 5 | Process qualified on similar space environment |
| Main electronics | | |
| – Electronic boards | 6 | Qualified on previous space environment (MIRI/JWST) |
| – Compressor process | 9 | Space qualified |
| – Space wire links | 9 | Space qualified |

Amici prisms, providing quite constant low dispersion across the selected wavelength ranges. The adopted scheme ensures that cross-talk due to both diffraction (the incoherent cross talk) and interference effects (the coherent cross talk) are well below $10^{-2}$. The main specifications listed in Table 4 were derived from the main equation governing this design.

3.8 Science detectors

Two large $4 \times 4$ k CCDs are implemented in the IFS for scientific observations. The selected elements are E2V detectors CCD 231-84 back illuminated with high TRL requiring thermal regulation. Table 5 shows their characteristics.

3.9 TRL summary

Table 6 draws a summary of the TRLs relevant to the SPICES mission.

## 4 Pointing issues

The pointing requirement is an important aspect of the SPICES concept. At the instrument level, the alignment of the star onto the coronagraph must be very good to avoid stellar leakage and achieve the largest contrast. From simulations, we obtained a requirement of 0.1 mas at the level of the coronagraph over a timescale that is representative of the instrument stability. Although observing time can be several days, the wavefront correction will be certainly re-actualized at a higher frequency, typically between a few hours and a day (a correct estimation will depend on the final architecture). In addition, a telescope pointing of typically 2–10 mas is required to avoid beam walk on the primary and secondary, which would otherwise cause an unacceptable decorrelation of the wavefront over time. The demanding accuracy necessitates a specific strategy. We propose a three-stage procedure (see Table 7 for summary).

First, a coarse pointing is achieved by the spacecraft service module, within an accuracy of a few arcseconds. This performance is achieved on European satellites like GAIA or HERSCHEL. Attitude measurement is performed by gyro-stellar hybridization, and a cold gas µ-propulsion system is used as an actuator to reduce vibration unlike reaction wheels.

As on GAIA, the second stage makes use of the payload itself as an attitude sensor. Pointing accuracy as required can be achieved with respect to the target star direction: some mas pointing, and some mas/s pointing stability. The star measurement by the payload is obtained directly on the target star therefore limiting the differential effects. Two potential systems are considered: either the use of light reflected by the filters as shown in the conceptual design (Fig. 8) or the use of the starlight rejected by the coronagraph at the Lyot stop. Although the first solution has the advantage of saving all the light unused by the science channels, the second concept has been simulated numerically at a level of 0.05 mas. One part of the error signal generated by the tip tilt sensor is off-loaded to the spacecraft attitude control system (ACS) and the telescope to allow a pointing precision of typically 10 mas. For comparison, JWST performs small angle maneuver of <2" with a 10–20 mas accuracy [33]. An optimized strategy with a monolithic telescope is certainly able to achieve a better precision. However, a dedicated system might be needed to reach 1 mas if needed.

**Table 7** Summary of pointing requirements for SPICES

|  | Spacecraft | Telescope | Coronagraph |
|---|---|---|---|
| 1st stage | >10 arcsec |  |  |
| 2nd stage | 10–100 mas | 2–10 mas |  |
| 3rd stage |  |  | 0.1 mas |

The second part of the error signal on the pointing detector is sent to an internal tip tilt corrector and provides the third stage of pointing. The usual way to correct for small tip tilt variations is to make use of a Fine Steering Mirror. A preferable solution for SPICES is to use a tip/tilt mount like on ground-based telescope, so that tip/tilt and higher order corrections are obtained by the same element in the same optical plane (DM is on a tip/tilt mount). The ACCESS proposal [44] demonstrates the ability of the telescope ACS to achieve the pointing specifications (1 mas for the telescope and 0.1 mas at the coronagraph). We refer to this public document for details.

## 5 Performance assessment

The performance results presented in the ESA proposal were preliminary but a more detailed study will be published soon. Here, we illustrate the expected performance of SPICES in terms of characterization for only two cases of planets, a giant and a super-Earth. A numerical model of the instrument was built to assess the performance of SPICES. We did not simulate the actual layout presented in Fig. 10 but only the main functionalities shown in Fig. 8. We assume some realistic level of static WFE (20 nm rms), a $64 \times 64$ DM, a Vector Vortex Coronagraph (achromatic), and the use of a self-coherent camera (SCC). The simulation iterates on the wavelength (40 channels with the resolution ranging from 35 to 80) to produce xyλ data cubes as the final product of the instrument (we did not account for polarimetry at this stage). We consider that the SCC provides a perfect estimation of the phase. The SCC measures the wavefront in the focal plane, drives the DM, and finally improves the image quality by post-processing. A random phase error of 0.5 nm rms is added at each spectral channel to account for chromaticity. The sampling is three pixels at the shortest wavelength. We also assume imperfection of flat field (1%) and read-out-noise of 1e-/pix with a full-well capacity of 300 ke-.

Figure 11 shows the contrast curves without noises obtained with a setup combining a DM and a coronagraph (blue curves) and the same setup plus a SCC postprocessing (red curves) that improves the speckle rejection by one order of magnitude in some spectral channels. We satisfy the requirement of a few $10^{-9}$ contrast at 2 $\lambda$/D. To illustrate the performance we consider two model spectra of planets. We assume a G2V type for the host star and a global throughput of 16%. We account for the photon noise, RON and flat field noise. The exposure time is determined so that the median SNR on the spectrum is 10. The left panel in Fig. 12 compares the measured spectrum (red curve) to the theoretical one (black curve) for a Jupiter (1 $R_J$) at 2 AU at 10 pc (models from [38]). The spectrum mainly shows methane features that are well recovered in 15 h (Table 8). The right panel stands for a super-Earth (2.5 $R_⊕$) at 1 AU at 6 pc [37]. The absorption lines are due to water, dioxygen, and ozone. Here again the main features are recovered but for a longer exposure time of 110 h.

**Fig. 11** Simulated 1-sigma contrast curves of SPICES for each spectral channel between 450 and 900 nm

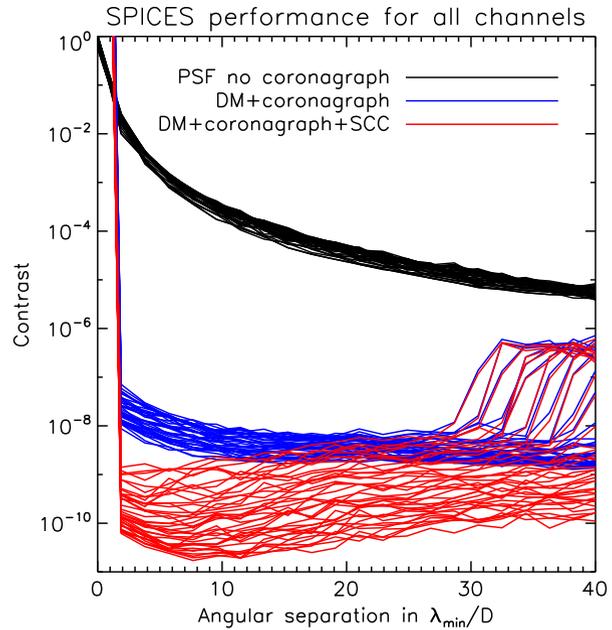

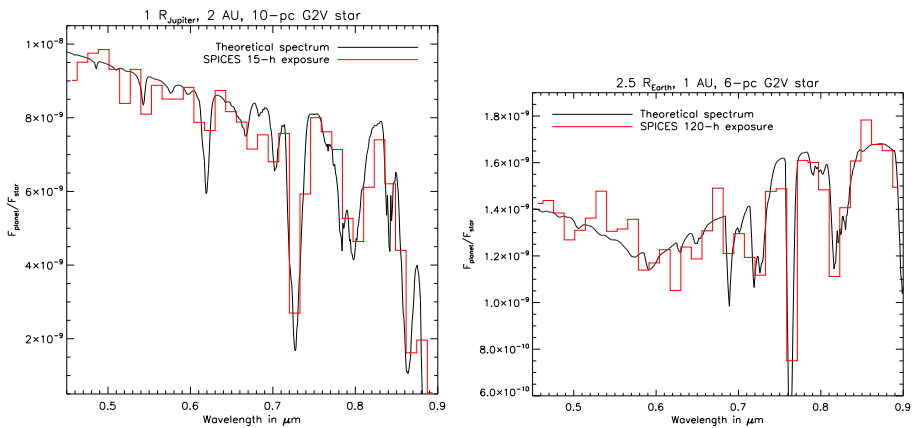

**Fig. 12** Contrast spectra of a Jupiter-like planet (*left*) and a Super-Earth (*right*) measured by SPICES and compared to the theoretical ones. The main molecular lines are detectable at: 0.62, 0.74, 0.79, and 0.86 $\mu$m ($CH_4$) for the giant planet and 0.6 ($O_3$), 0.72 ($H_2O$), 0.76 ($O_2$), and 0.82 ($H_2O$) $\mu$m for the telluric one

**Table 8** Signal to noise ratio corresponding to spectra in Fig. 12

| Jupiter at 2 AU | | Super Earth at 1 AU | |
|---|---|---|---|
| Feature | SNR | Feature | SNR |
| Continuum | 14.3 (mean) | Continuum | 12.2 (mean) |
| $CH_4$ band @ 0.62 μm | 14.4 | $O_3$ line @ 0.6 μm | 10.6 |
| $CH_4$ band @ 0.74 μm | 6.1 | $H_2O$ line @ 0.72 μm | 7.6 |
| $CH_4$ band @ 0.79 μm | 11.7 | $O_2$ line @ 0.76 μm | 8.0 |
| $CH_4$ band @ 0.86 μm | 2.3 | $H_2O$ line @ 0.82 μm | 7.4 |

In the high contrast regime delivered by SPICES, the detection is essentially photon noise limited. Therefore, we use the previous results to scale exposure times according to distance and spectral type of the star. The exposure time for a given SNR and flux is proportional to the square of the distance. For a given SNR and star distance, it is inversely proportional to the flux. Figure 13 presents the results of this study for the Jupiter planet (left) and the super-Earth (right). For example, a Jupiter planet at 4 AU is resolved by SPICES at 20 pc and requires an exposure time 4 times greater than the one at 2 AU. If it orbits a F0V star, the required exposure time is 60 h. These analytical relations are now being tested by simulations and will be presented in a forthcoming paper. This preliminary analysis tells us that SPICES will be most efficient for

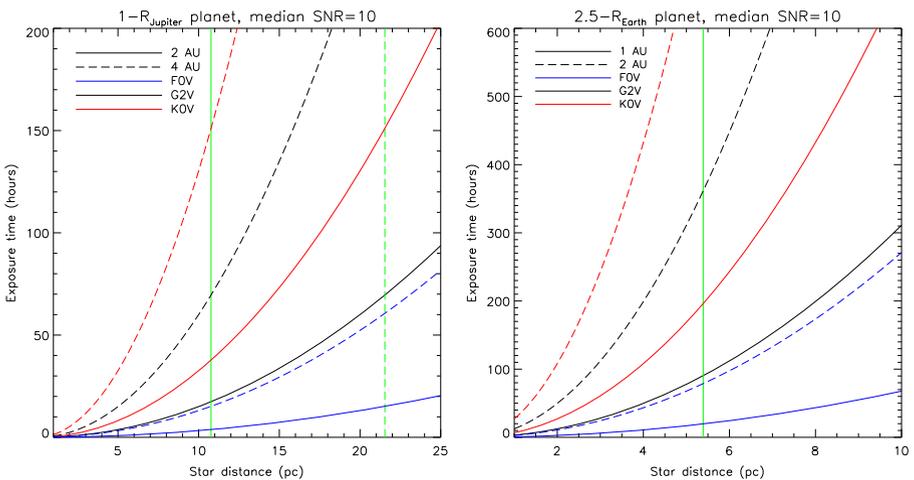

**Fig. 13** Evolution of the exposure time, for a median SNR, of 10 as a function of the star distance and spectral type for a Jupiter-like planet at 2 and 4 AU (*left*) and a Super-Earth at 1 and 2 AU (*right*). The *vertical green lines* indicate the maximum star distance where the planet is resolved by SPICES (given that IWA $= 2\lambda/D$) maneuvers

giant planets in the 1–4 AU range, and that tens of hours are necessary to achieve good SNR.

## 6 Operations concept

SPICES is a pointing mission operated by way of autonomous execution and work plan periodically up-loaded from the ground. A large majority of targets, those in the core program, will be selected before the launch. The mission will consist of a series of long observations where the telescope will stare at the same target for typically a few days. As explained above, a full observation on a target requires four different setups of the instrument where the polarization states are flipped by means of a modulator. Before an observation starts, the instrument will have to measure the phase aberrations to drive the DM in an iterative way. Since the instrumental environment is likely to change over time, this measurement will be repeated several times in a sequence. Once the phase is measured, the correction is applied to the DM, which is maintained in this position for some hours (the exact frequency will depend on the overall stability). Minimum individual exposures will be determined by the detector readout time (about 10 s) and maximum exposures by the cosmic rays (a few hundred seconds). Only averaged sequences of 30 min will be transferred to the Earth. After the full sequence is completed, the telescope will slew to the next target and the data will transfer during the re-pointing.

## 7 Spacecraft concept

The spacecraft is three-axes stabilized and controlled using a coarse stage for large maneuvers and a fine stage for observation. The fine stage shall be able to point the spacecraft to the target and to stabilize the line of sight below 10 mas PtV, with information provided by the payload itself.

The telescope core structure is constituted by a nearly SiC L-shape frame which provides easy fixation interfaces to the primary and secondary mirrors, the focal planes and optical items of the polarimeters and coronagraphs. In this way, the complete optical path is then fixed to this main structural frame (Fig. 14).

Apart from the deformable mirrors, which require a room temperature (10–25°C) working point, the unique specific thermal requirement for the rest of the payload is the stability needed during acquisition. Telescope European manufacturing techniques based on SiC can preserve optical surface qualities at low temperature. Assuming that the thermal flux in L2 is not sufficient to easily ensure a global warming of the payload, the thermal concept is based on a 200 K passive cooling, ensuring a high stability due to the very high thermal inertia and a passive equilibrium with the environment. The space telescope GAIA, having very similar requirements to SPICES, is based on the same concept and is passively stabilized at a few mK. A dedicated warming system will be used for the deformable mirrors. Each mirror will be thermally isolated from the rest of the payload in order to avoid thermal leaks. We consider the possibility of encapsulating the DMs in a chamber with a BK7 window in order to prevent radiative heating from the DM to the rest of the payload and allow stabilization at the mK.

**Fig. 14** Opto-mechanical views of the telescope with the payload showing the L shape base plate on which the optical elements are fixed. The dimension of the telescope is 3,350 × 1,850 × 1,540 mm

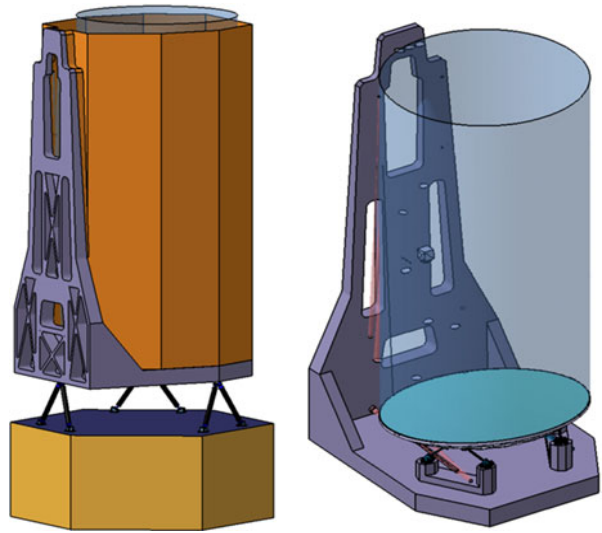

## 8 Conclusion

The SPICES concept presented here is an evolution of the former SEE COAST proposal to Cosmic Vision 2007 [36] in which several aspects of the mission have been significantly improved: availability of targets, conceptual design of the instrument, preliminary optical design, evaluation of risks, identification of subsystems with high TRL and technological development plans.

SPICES was not selected by ESA in 2010 but several technological developments related to the mission are being pursued in our institutes, especially concerning coronagraphy and wavefront correction. Some efforts are still needed to demonstrate (at the TRL 5 level) the ability to reach $\sim 10^{-9}$ contrast over large bandwidths ($\sim$50%) with the VVC, as well as the fine measurement and correction of the aberrations with the SCC.

The landscape of exoplanet science for the next decade strongly requires the availability of a mission like SPICES to study the atmosphere of long-period planets (that we expect to be comparable to those of giants in the Solar System) in order to complement other direct detection programs, especially the spectroscopy of close-in transiting planets and direct imaging of warm giant planets.

**Acknowledgements** We would like to thanks H. Boithias, E. Sein at ASTRIUM France for their technical support and A. Laurens, O. Lamarle at CNES for both financial and technical support during this ESA Cosmic Vision proposal study. We are also grateful to the referee for relevant comments which helped to improve this paper.

# References

1. Antichi, J., Dohlen, K., Gratton, R.G., Mesa, D., Claudi, R.U., Giro, E., Boccaletti, A., Mouillet, D., Puget, P., Beuzit, J.: BIGRE: a low cross-talk integral field unit tailored for extrasolar planets imaging spectroscopy. Astrophys. J. **695**, 1042–1057 (2009)
2. Arnold, L., Schneider, J.: The detectability of extrasolar planet surroundings. I. Reflected-light photometry of unresolved rings. Astron. Astrophys. **420**, 1153–1162 (2004)
3. Baraffe, I., Chabrier, G., Barman, T.S., Allard, F., Hauschildt, P.H.: Evolutionary models for cool brown dwarfs and extrasolar giant planets. The case of HD 209458. Astron. Astrophys. **402**, 701–712 (2003)
4. Barman, T.S., Macintosh, B., Konopacky, Q.M., Marois, C.: Clouds and chemistry in the atmosphere of extrasolar planet HR8799b. Astrophys. J. **733**, 65–83 (2011)
5. Baudoz, P., Boccaletti, A., Baudrand, J., Rouan, D.: The self-coherent camera: a new tool for planet detection. In: Aime, C., Vakili, F. (eds.) IAU Colloq. 200: Direct Imaging of Exoplanets: Science & Techniques, pp. 553–558 (2006)
6. Baudoz, P., Boccaletti, A., Rouan, D.: Multiple-stage four quadrant phase mask coronagraph. In: In the Spirit of Bernard Lyot: the Direct Detection of Planets and Circumstellar Disks in the 21st Century (2007)
7. Belikov, R., Pluzhnik, E., Connelley, M.S., Witteborn, F.C., Greene, T.P., Lynch, D.H., Zell, P.T., Guyon, O.: Laboratory demonstration of high-contrast imaging at 2 λ/D on a temperature-stabilized testbed in air. In: Society of Photo-Optical Instrumentation Engineers (SPIE) Conference Series, Society of Photo-Optical Instrumentation Engineers (SPIE) Conference Series, vol. 7731 (2010)
8. Beuzit, J., Feldt, M., Dohlen, K., Mouillet, D., Puget, P., Wildi, F., Abe, L., Antichi, J., Baruffolo, A., Baudoz, P., Boccaletti, A., Carbillet, M., Charton, J., Claudi, R., Downing, M., Fabron, C., Feautrier, P., Fedrigo, E., Fusco, T., Gach, J., Gratton, R., Henning, T., Hubin, N., Joos, F., Kasper, M., Langlois, M., Lenzen, R., Moutou, C., Pavlov, A., Petit, C., Pragt, J., Rabou, P., Rigal, F., Roelfsema, R., Rousset, G., Saisse, M., Schmid, H., Stadler, E., Thalmann, C., Turatto, M., Udry, S., Vakili, F., Waters, R.: SPHERE: a planet finder instrument for the VLT. In: Society of Photo-Optical Instrumentation Engineers (SPIE) Conference Series, vol. 7014 (2008)
9. Boccaletti, A., Abe, L., Baudrand, J., Daban, J., Douet, R., Guerri, G., Robbe-Dubois, S., Bendjoya, P., Dohlen, K., Mawet, D.: Prototyping coronagraphs for exoplanet characterization with SPHERE. In: Society of Photo-Optical Instrumentation Engineers (SPIE) Conference Series, vol. 7015 (2008)
10. Boccaletti, A., Riaud, P., Baudoz, P., Baudrand, J., Rouan, D., Gratadour, D., Lacombe, F., Lagrange, A.: The four-quadrant phase mask coronagraph. IV. First light at the very large telescope. Publ. Astron. Soc. Pac. **116**, 1061–1071 (2004)
11. Bowler, B.P., Liu, M.C., Dupuy, T.J., Cushing, M.C.: Near-infrared spectroscopy of the extrasolar planet hr 8799 b. Astrophys. J. **723**, 850–868 (2010)
12. Buenzli, E., Thalmann, C., Vigan, A., Boccaletti, A., Chauvin, G., Augereau, J.C., Meyer, M.R., Ménard, F., Desidera, S., Messina, S., Henning, T., Carson, J., Montagnier, G., Beuzit, J.L., Bonavita, M., Eggenberger, A., Lagrange, A.M., Mesa, D., Mouillet, D., Quanz, S.P.: Dissecting the moth: discovery of an off-centered ring in the HD 61005 debris disk with high-resolution imaging. Astron. Astrophys. **524**, L1–5 (2010)
13. Cahoy, K.L., Marley, M.S., Fortney, J.J.: Exoplanet albedo spectra and colors as a function of planet phase, separation, and metallicity. Astrophys. J. **724**, 189–214 (2010)
14. Casertano, S., Lattanzi, M.G., Sozzetti, A., Spagna, A., Jancart, S., Morbidelli, R., Pannunzio, R., Pourbaix, D., Queloz, D.: Double-blind test program for astrometric planet detection with Gaia. Astron. Astrophys. **482**, 699–729 (2008)

15. Catala, C., Appourchaux, T., Plato Mission Consortium: PLATO: PLAnetary transits and oscillations of stars. J. Phys. Conf. Ser. **271**(012084), 1–7 (2011)
16. Chauvin, G., Lagrange, A., Dumas, C., Zuckerman, B., Mouillet, D., Song, I., Beuzit, J., Lowrance, P.: A giant planet candidate near a young brown dwarf. Direct VLT/NACO observations using IR wavefront sensing. Astron. Astrophys. **425**, L29–L32 (2004)
17. Claudi, R.U., Turatto, M., Giro, E., Mesa, D., Anselmi, U., Bruno, P., Cascone, E., de Caprio, V., Desidera, S., Dorn, R., Fantinel, D., Finger, G., Gratton, R.G., Lessio, L., Lizon, J.L., Salasnic, B., Scuderi, S., Dohlen, K., Beuzit, J.L., Puget, P., Antichi, J., Hubin, N., Kasper, M.: SPHERE IFS: the spectro differential imager of the VLT for exoplanets search. In: Society of Photo-Optical Instrumentation Engineers (SPIE) Conference Series, vol. 7735 (2010)
18. Galicher, R., Baudoz, P., Rousset, G.: Wavefront error correction and Earth-like planet detection by a self-coherent camera in space. Astron. Astrophys. **488**, L9–L12 (2008)
19. Galicher, R., Baudoz, P., Rousset, G., Totems, J., Mas, M.: Self-coherent camera as a focal plane wavefront sensor: simulations. Astron. Astrophys. **509**, 31–45 (2010)
20. Give'on, A., Kern, B., Shaklan, S., Moody, D.C., Pueyo, L.: Broadband wavefront correction algorithm for high-contrast imaging systems. In: Society of Photo-Optical Instrumentation Engineers (SPIE) Conference Series, vol. 6691 (2007)
21. Howard, A.W., Marcy, G.W., Johnson, J.A., Fischer, D.A., Wright, J.T., Isaacson, H., Valenti, J.A., Anderson, J., Lin, D.N.C., Ida, S.: The occurrence and mass distribution of close-in super-earths, Neptunes, and Jupiters. Science **330**, 653–655 (2010)
22. Kalas, P., Graham, J.R., Chiang, E., Fitzgerald, M.P., Clampin, M., Kite, E.S., Stapelfeldt, K., Marois, C., Krist, J.: Optical images of an exosolar planet 25 light-years from earth. Science **322**, 1345–1348 (2008)
23. Karkoschka, E.: Spectrophotometry of the Jovian planets and Titan at 300- to 1,000-nm wavelength: the methane spectrum. Icarus **111**, 174–192 (1994)
24. Kraus, A.L., Ireland, M.J.: LkCa 15: A Young Exoplanet Caught at Formation? (2011). arXiv:1110.3808v1
25. Lagrange, A., Gratadour, D., Chauvin, G., Fusco, T., Ehrenreich, D., Mouillet, D., Rousset, G., Rouan, D., Allard, F., Gendron, É., Charton, J., Mugnier, L., Rabou, P., Montri, J., Lacombe, F.: A probable giant planet imaged in the β Pictoris disk. VLT/NaCo deep L'-band imaging. Astron. Astrophys. **493**, L21–L25 (2009)
26. Levine, M., Lisman, D., Shaklan, S., Kasting, J., Traub, W., Alexander, J., Angel, R., Blaurock, C., Brown, M., Brown, R., Burrows, C., Clampin, M., Cohen, E., Content, D., Dewell, L., Dumont, P., Egerman, R., Ferguson, H., Ford, V., Greene, J., Guyon, O., Hammel, H., Heap, S., Ho, T., Horner, S., Hunyadi, S., Irish, S., Jackson, C., Kasdin, J., Kissil, A., Krim, M., Kuchner, M., Kwack, E., Lillie, C., Lin, D., Liu, A., Marchen, L., Marley, M., Meadows, V., Mosier, G., Mouroulis, P., Noecker, M., Ohl, R., Oppenheimer, B., Pitman, J., Ridgway, S., Sabatke, E., Seager, S., Shao, M., Smith, A., Soummer, R., Stapelfeldt, K., Tenerell, D., Trauger, J., Vanderbei, R.: Terrestrial Planet Finder Coronagraph (TPF-C) Flight Baseline Concept. ArXiv e-prints (2009)
27. Lovis, C., Ségransan, D., Mayor, M., Udry, S., Benz, W., Bertaux, J.L., Bouchy, F., Correia, A.C.M., Laskar, J., Lo Curto, G., Mordasini, C., Pepe, F., Queloz, D., Santos, N.C.: The HARPS search for southern extra-solar planets. XXVIII. Up to seven planets orbiting HD 10180: probing the architecture of low-mass planetary systems. Astron. Astrophys. **528**, 112–128 (2011)
28. Macintosh, B.A., Graham, J.R., Palmer, D.W., Doyon, R., Dunn, J., Gavel, D.T., Larkin, J., Oppenheimer, B., Saddlemyer, L., Sivaramakrishnan, A., Wallace, J.K., Bauman, B., Erickson, D.A., Marois, C., Poyneer, L.A., Soummer, R.: The gemini planet imager: from science to design to construction. In: Society of Photo-Optical Instrumentation Engineers (SPIE) Conference Series, vol. 7015 (2008)
29. Marois, C., Macintosh, B., Barman, T., Zuckerman, B., Song, I., Patience, J., Lafrenière, D., Doyon, R.: Direct imaging of multiple planets orbiting the star HR 8799. Science **322**, 1348 (2008)
30. Mawet, D., Riaud, P., Baudrand, J., Baudoz, P., Boccaletti, A., Dupuis, O., Rouan, D.: The four-quadrant phase-mask coronagraph: white light laboratory results with an achromatic device. Astron. Astrophys. **448**, 801–808 (2006)
31. Mawet, D., Serabyn, E., Stapelfeldt, K., Crepp, J.: Imaging the debris disk of HD 32297 with a phase-mask coronagraph at high strehl ratio. Astrophys. J. Lett. **702**, L47–L50 (2009)

32. Mawet, D., Trauger, J.T., Serabyn, E., Moody Jr., D.C., Liewer, K.M., Krist, J.E., Shemo, D.M., O'Brien, N.A.: Vector vortex coronagraph: first results in the visible. In: Society of Photo-Optical Instrumentation Engineers (SPIE) Conference Series, vol. 7440 (2009)

33. Nelan, E.: JWST science instrument target acquisition concepts: JWST-STScI-000405. Tech. rep., STScI (2005)

34. Rouan, D., Riaud, P., Boccaletti, A., Clénet, Y., Labeyrie, A.: The four-quadrant phase-mask coronagraph. I. Principle. Publ. Astron. Soc. Pac. **112**, 1479–1486 (2000)

35. Schmid, H.M., Beuzit, J.L., Mouillet, D., Waters, R., Buenzli, E., Boccaletti, A., Dohlen, K., Feldt, M., SPHERE Consortium: Polarimetry of extra-solar planets and circumstellar disks with ZIMPOL/SPHERE. In: In the Spirit of Lyot 2010 (2010)

36. Schneider, J., Boccaletti, A., Mawet, D., Baudoz, P., Beuzit, J.L., Doyon, R., Marley, M., Stam, D., Tinetti, G., Traub, W., Trauger, J., Aylward, A., Cho, J.Y.K., Keller, C.U., Udry, S., SEE-COAST team: super earth explorer: a coronagraphic off-axis space telescope. Exp. Astron. **23**, 357–377 (2009)

37. Stam, D.M.: Spectropolarimetric signatures of Earth-like extrasolar planets. Astron. Astrophys. **482**, 989–1007 (2008)

38. Stam, D.M., Hovenier, J.W., Waters, L.B.F.M.: Using polarimetry to detect and characterize Jupiter-like extrasolar planets. Astron. Astrophys. **428**, 663–672 (2004)

39. Swain, M.R., Tinetti, G., Vasisht, G., Deroo, P., Griffith, C., Bouwman, J., Chen, P., Yung, Y., Burrows, A., Brown, L.R., Matthews, J., Rowe, J.F., Kuschnig, R., Angerhausen, D.: Water, methane, and carbon dioxide present in the dayside spectrum of the exoplanet HD 209458b. Astrophys. J. **704**, 1616–1621 (2009)

40. Swain, M.R., Vasisht, G., Tinetti, G., Bouwman, J., Chen, P., Yung, Y., Deming, D., Deroo, P.: Molecular signatures in the near-infrared dayside spectrum of HD 189733b. Astrophys. J. Lett. **690**, L114–L117 (2009)

41. Tamura, M., Hodapp, K., Takami, H., Abe, L., Suto, H., Guyon, O., Jacobson, S., Kandori, R., Morino, J., Murakami, N., Stahlberger, V., Suzuki, R., Tavrov, A., Yamada, H., Nishikawa, J., Ukita, N., Hashimoto, J., Izumiura, H., Hayashi, M., Nakajima, T., Nishimura, T.: Concept and science of HiCIAO: high contrast instrument for the Subaru next generation adaptive optics. In: Society of Photo-Optical Instrumentation Engineers (SPIE) Conference Series, vol. 6269 (2006)

42. Thalmann, C., Grady, C.A., Goto, M., Wisniewski, J.P., Janson, M., Henning, T., Fukagawa, M., Honda, M., Mulders, G.D., Min, M., Moro-Martín, A., McElwain, M.W., Hodapp, K.W., Carson, J., Abe, L., Brandner, W., Egner, S., Feldt, M., Fukue, T., Golota, T., Guyon, O., Hashimoto, J., Hayano, Y., Hayashi, M., Hayashi, S., Ishii, M., Kandori, R., Knapp, G.R., Kudo, T., Kusakabe, N., Kuzuhara, M., Matsuo, T., Miyama, S., Morino, J.I., Nishimura, T., Pyo, T.S., Serabyn, E., Shibai, H., Suto, H., Suzuki, R., Takami, M., Takato, N., Terada, H., Tomono, D., Turner, E.L., Watanabe, M., Yamada, T., Takami, H., Usuda, T., Tamura, M.: Imaging of a transitional disk gap in reflected light: indications of planet formation around the young solar analog LkCa 15. Astrophys. J. Lett. **718**, L87–L91 (2010)

43. Trauger, J., Give'on, A., Gordon, B., Kern, B., Kuhnert, A., Moody, D., Niessner, A., Shi, F., Wilson, D., Burrows, C.: Laboratory demonstrations of high-contrast imaging for space coronagraphy. In: Society of Photo-Optical Instrumentation Engineers (SPIE) Conference Series, vol. 6693 (2007)

44. Trauger, J.T., et al.: ACCESS: a space coronagraph concept for direct imaging and spectroscopy of exoplanetary systems. Tech. rep., NASA-JPL (2009). http://exep.jpl.nasa.gov/files/exep/ACCESS_Compiled_Report_Public_090610.pdf

45. Trauger, J.T., Traub, W.A.: A laboratory demonstration of the capability to image an Earth-like extrasolar planet. Nature **446**, 771–773 (2007)

46. Wittenmyer, R.A., Tinney, C.G., O'Toole, S.J., Jones, H.R.A., Butler, R.P., Carter, B.D., Bailey, J.: On the frequency of jupiter analogs. Astrophys. J. **727**, 102–118 (2011)